# Symbolic landforms created by ancient earthworks near Lake Titicaca

**Amelia Carolina Sparavigna**
Dipartimento di Fisica, Politecnico di Torino
C.so Duca degli Abruzzi 24, Torino, Italy

**Abstract**
Interesting landforms created by an ancient network of earthworks are shown, using Google satellite imagery enhanced by an image processing. This network covers a large part of the land near the Titicaca Lake. Satellite images clearly display the slopes of hills criss-crossed with terrace walls and the surfaces of the plains covered with raised fields, indicating that this was once a highly productive agricultural place for the south central Andes. Some of the landforms are rather remarkable, having a clear symbolic function. Among them, there are structures which seem to represent birds, where ponds are their eyes.

Any landform composed of fine-grained materials evolves in wide and flat relieves, due to the down-slope transport of its materials over time. Earthworks, which are artificial landforms, are subjected to the same destiny, to be widened and flattened as a consequence of the natural degradation processes [1,2]. The evolution of ancient earthworks indicates that a variety of initial earthwork forms can result in a sinusoidal profile apparent on the current landscape [3]. Therefore, these ancient structures can be clearly displayed by satellite imagery as a texture superimposed to the background landform. In some cases, they remain quite visible and, on them, the modern structures.

In this paper, a wonderful example of a landform created by a really huge network of earthworks is discussed, using Google satellite Maps: this network covers a total of 120,000 hectares of the land near the Titicaca Lake, being the result of an almost unimaginable agricultural effort of ancient Andean people. In fact, some of the landforms near Lake Titicaca are rather remarkable, having a clear symbolic meaning, and can be considered as geo-glyphs. Some represent birds, where circular ponds are their eyes. Other animals can be observed in a complete survey of the lands near the lake.

The network of earthworks is the remains of an extensive ancient agricultural system built and used by Andean peoples centuries ago, throughout the vast high plain surrounding Titicaca. People created a system of raised fields, which were large elevated planting platforms, with the corresponding drainage canals. This system improved soil conditions, the temperature and moisture conditions for crops. The remains of prehistoric raised fields are then providing evidence of the impressive engineering abilities of the peoples who lived there in pre-Columbian times. Moreover, this finding contradicts the opinion that considers the lands of the Lake Titicaca to be unproductive agriculturally. Archaeology and the satellite imagery, demonstrate the past richness of the area, due to this vast complex of agricultural earthworks.

The local farmers call the artificial landforms "waru waru" or "camellones" (pre-Hispanic raised fields are present in other regions too [4-7]). The local farmers of Titicaca had no idea that these textures are the persisting evidence of remarkable skills of their ancestors, until 1981, when Clark Erickson, University of Illinois, recognized the significance of waru waru. He and other researchers started an experimental reintroduction of raised fields, in the Huatta, a land near the lake, in Peru. They persuaded some local farmers to rebuild a few of the raised fields, plant them in indigenous crops, and farm in traditional manner. Archaeological and experimental data suggest that raised fields might be more appropriate for the region [7].

Let us observe the satellite images. Lake Titicaca sits 3,811 m above sea level, in a basin high in the Andes on the border of Peru and Bolivia. The western part of the lake lies within the Puno Region of Peru, and the eastern side is located in the Bolivian La Paz Department. Both regions have the slopes of the hills criss-crossed with terrace walls. Some parts of the plain surfaces are covered with raised fields, indicating that this was once a highly productive agricultural place for the south central Andes. As earthworks, raised fields are constructed by excavating parallel canals and piling the earth between them creating long and low mounds, surfaces being flat or convex [7]. These raised platforms created a local micro-environment, able to reduce the frost risk for crops. The canals between raised fields act as sources of moisture during the periods of drought. Moreover, water in the deep canals might have been used to cultivate aquatic plants and fish, as well as attract lake birds [7].

The raised fields of Titicaca have different forms and size, generally being 4-10 m wide, 10 to 100 m long, and 1 m tall. Some early fields were narrow ridges of 5 m wavelength. At a later time, the wavelength increased for larger fields to 10 m [7]. In spite of erosion, the network of these not so-high earthworks is clearly visible from the space. For instance, Figure 1 shows a piece of this land. Observing the figure, we can argue that the creation of these earthworks was previously planned, following the natural slope of the terrain.

Many other interesting drawings are displayed by the satellite imagery. The author wants to show some of them, because of their symbolic evident planning. They seem geo-glyphs. In Figure 2 we see a bird, where a circular pond is the eye. In Figure 3, it looks like a condor being represented on the surface. For the three images, a processing method [8] was used that the edges of the network. In Fig.4, there is an animal that could be a hedgehog. Another artificial landform that could be a geo-glyph (a fish or a tortoise) is located at coordinates approx. -15.6464,-70.132. Landforms in Figures 1-4 are in Peru districts. In Bolivia we see a large area (approx. coordinates -16.4275,
-68.5822) where the raised fields have a different style. Here too, we see beautiful landforms, showing snakes, birds and other objects, not so easy to figure out. A rule of thumb: to find the figures, look for circular ponds, because sometimes they can be the eye of an animal. Figure 6 shows a snake and a bird in the Bolivian country.

In conclusion, the paper showed that the previously proposed image processing of natural landforms [8] can be applied to the study of artificial landforms, such as geo-glyphs. After processing, having the possibility to observe all the minute details of structures, a comparison of considered symbolic landforms with those of other regions is more easy [9,10]. A future work is devoted to a survey of all the Titicaca Lake region.

**Notes and references**

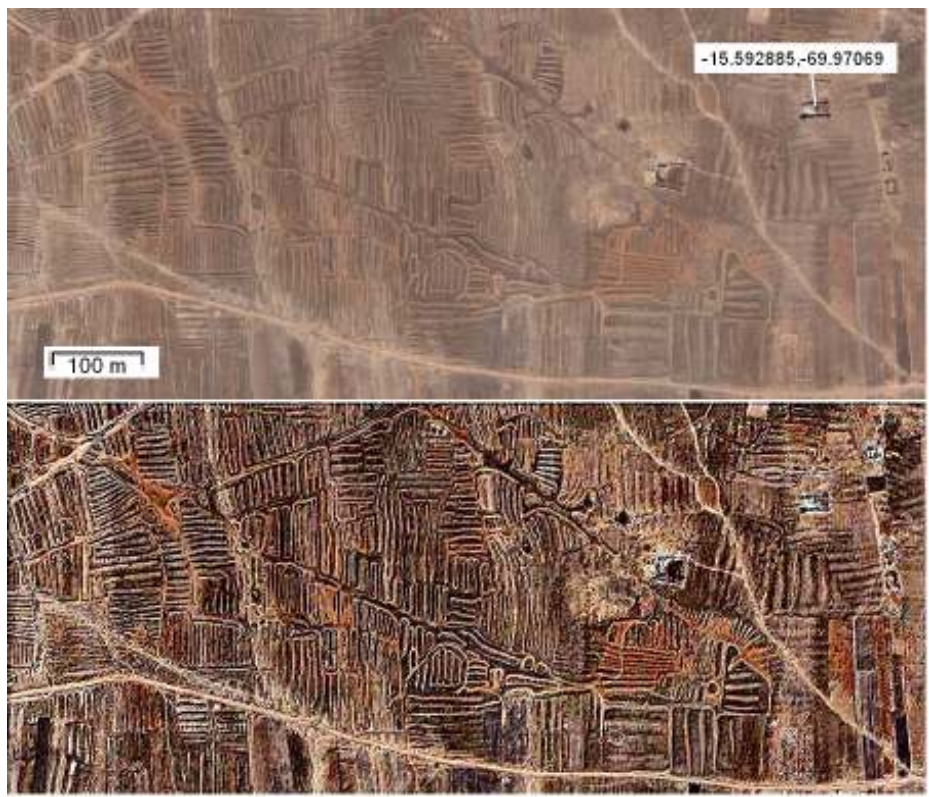

Figure 1: This is a part of the land near Lake Titicaca. In the upper panel, the original image from Google, in the lower one the image enhanced with a previously proposed method [8]. It looks like the head of a bird. In any case, we can argue that the creation of earthworks was previously planned, following the natural slope of the terrain. Note that the processing allows observing all the minute details.

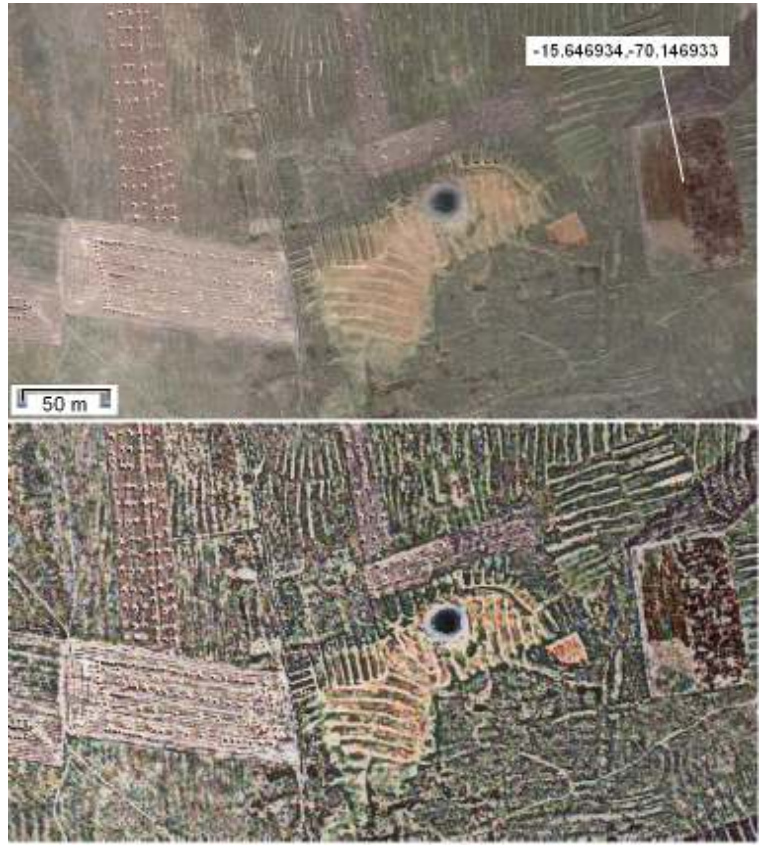

Figure 2: Many interesting drawings are displayed in the satellite imagery of this land. Among them, there are some which look as geo-glyphs. Here we see a bird, where a circular pond is the eye. In the upper panel, the original image from Google, in the lower part the image enhanced with a previously proposed method [8].

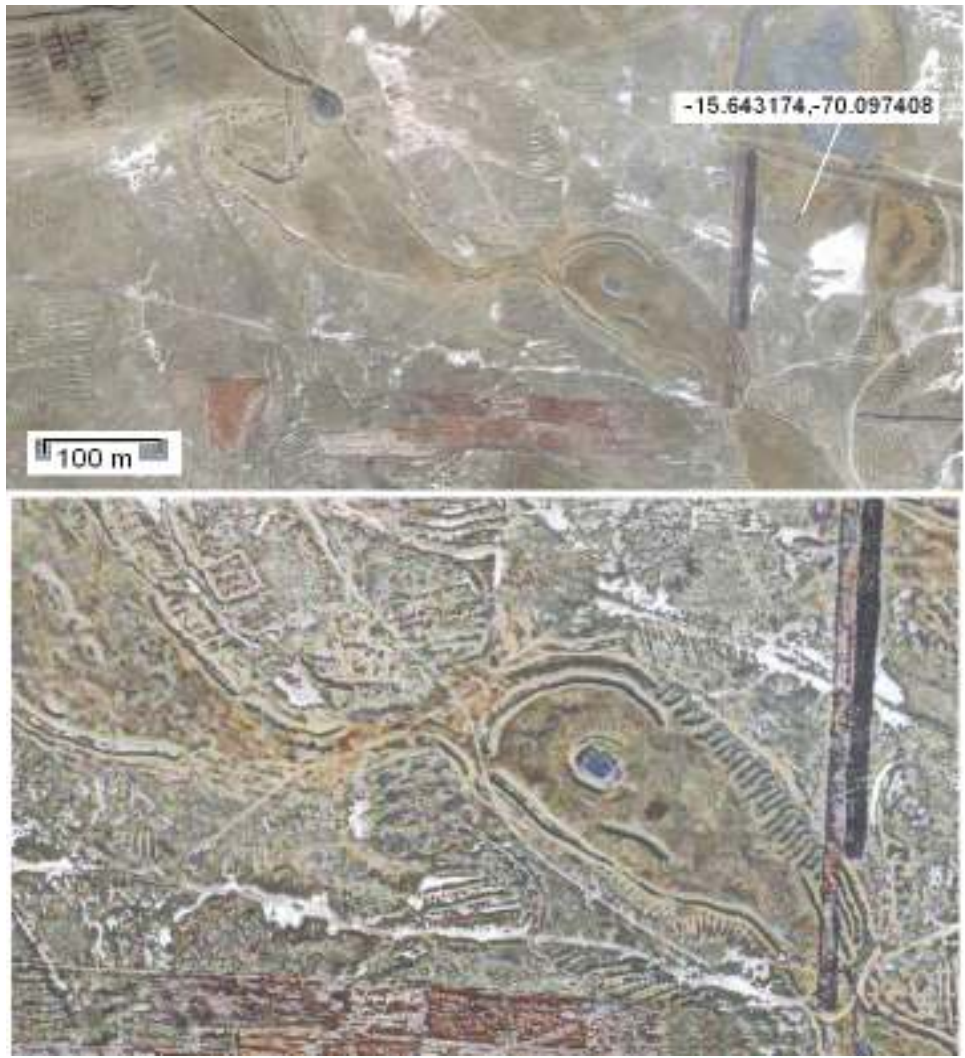

Figure 3. This landform appears as a geo-glyph representing a condor. In the upper panel, the original image from Google, in the lower part the image, the head enhanced with a previously proposed method [8].

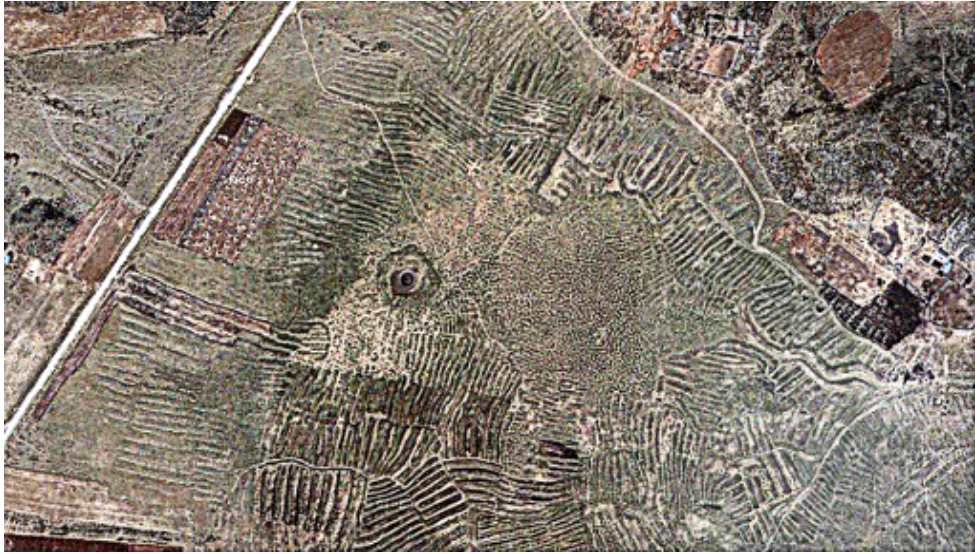

Figure 4: This artificial landform could represent a hedgehog. Coordinates are -15.65154,-70.1334. The figure shows an image obtained from a Google one, processed as in [8].

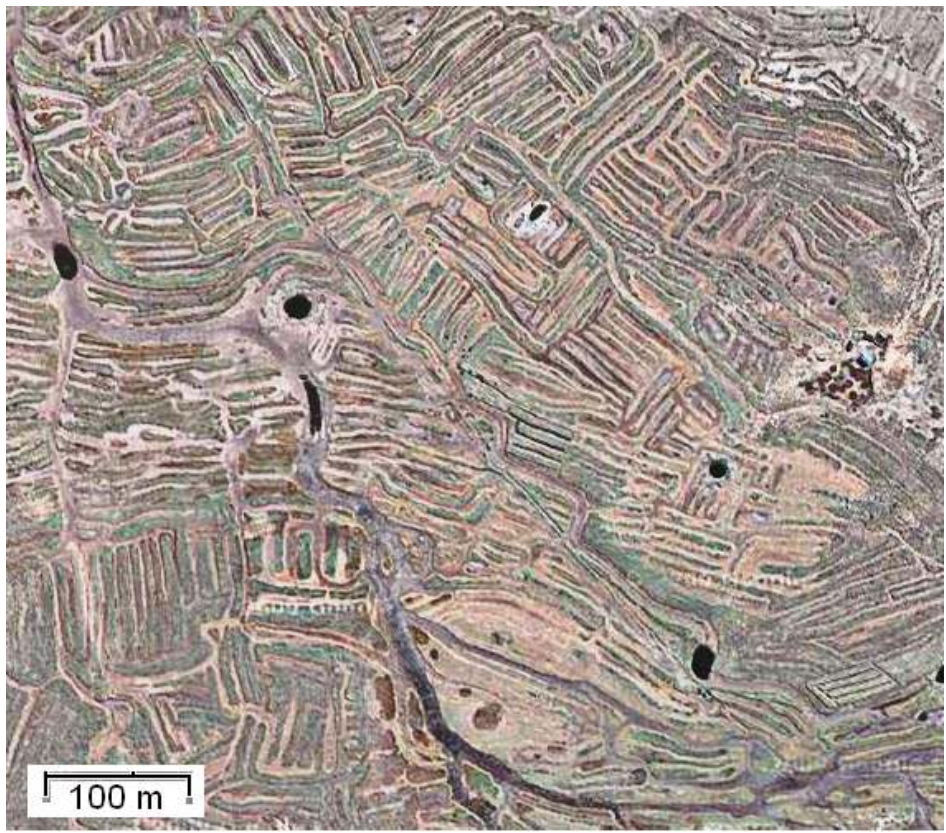

Figure 5: A snake and a bird in a plain region of Bolivia, near the lake. Form an image obtained by Google Maps, processed with a previously proposed method [8]. Note the different style of this landform.